\documentclass[runningheads]{llncs}

\usepackage[T1]{fontenc}
\usepackage{graphicx}
\usepackage{amsmath,amssymb,amsfonts}
\usepackage{algorithmic}
\usepackage{textcomp}
\usepackage{array}
\usepackage{multirow}
\usepackage{booktabs}
\usepackage{cite}
\usepackage{hyperref}
\usepackage{color}
\usepackage{bm} 

\usepackage[symbol]{footmisc}

\begin{document}

\title{MSMT-FN: Multi-segment Multi-task Fusion Network for Marketing Audio Classification}

\titlerunning{MSMT-FN for Marketing Audio Classification}

\author
{HongYu Liu\inst{1}$^{*}$ \and
Ruijie Wan\inst{1}$^{*}$ \and
Yueju Han\inst{1} \and
Junxin Li\inst{1} \and
Liuxing Lu\inst{1} \and 
Chao He\inst{1}$^{\dagger}$ \and
Lihua Cai\inst{1}$^{\dagger}$}

\institute{South China Normal University, Guangzhou, China\\
\email{hongyuliu,ruijiewan,yueju.han,junxinli,liuxing.lu,chaohe,lee.cai@m.scnu.edu.cn}
}
\authorrunning{Hongyu Liu et al.}
\renewcommand{\thefootnote}{\fnsymbol{footnote}}
\footnotetext[1]{These authors contributed equally.}

\renewcommand{\thefootnote}{\fnsymbol{footnote}}
\footnotetext[2]{Correspondent authors.}

\maketitle            

\begin{abstract}
Audio classification plays an essential role in sentiment analysis and emotion recognition, especially for analyzing customer attitudes in marketing phone calls. Efficiently categorizing customer purchasing propensity from large volumes of audio data remains challenging. In this work, we propose a novel Multi-Segment Multi-Task Fusion Network (MSMT-FN) that is uniquely designed for addressing this business demand. Evaluations conducted on our proprietary MarketCalls dataset, as well as established benchmarks (CMU-MOSI, CMU-MOSEI, and MELD), show MSMT-FN consistently outperforms or matches state-of-the-art methods. Additionally, our newly curated MarketCalls dataset will be available upon request, and the code base is made accessible at GitHub Repository MSMT-FN\footnote[3]{https://github.com/david188888/MSMT-FN}, to facilitate further research and advancements in audio classification domain.

\end{abstract}

\section{Introduction}
\label{sec:intro}
\vspace{-0.3cm}
Automatic voice-call robots have been increasingly adopted to replace human sales representatives to generate hundreds or even thousands of marketing phone calls each day for marketing purpose, immensely boosting marketing productivity. However, sales leads with high conversion potentials are invaluable, and existing manual classification approach incurs significant labor costs, making it hardly scalable.
A solution to efficiently and effectively classify these large-volume phone call recordings is urgently needed. 

The task of classifying purchase propensity in sales calls presents a unique set of challenges that distinguishes it from standard sentiment or emotion analysis tasks, such as those performed on the CMU-MOSI and CMU-MOSEI datasets. First, these are relatively longer duet conversations, rather than solo utterances. A customer's interest can evolve over the conversation — starting neutral to becoming interested in the product/service, or vice versa, which demands a model that can track the dynamics of intent over time. 
Second, the task requires decoding the subtlety of intent, which is different from sentiment. A customer can express polite and positive sentiment (e.g., "Thank you so much, that's very kind"), while firmly refuse the solicitation. A standard sentiment classifier would likely fail here, as purchase propensity is a more complex, goal-oriented construct. Third, the duet in the marketing call engage in the conversation in different approaches (i.e., the salesperson often follows a script while the potential customer reacts spontaneously). The most valuable signals of intent often lie in the customer's acoustic cues such as hesitation, tone, and engagement in response to the salesperson's pitch. Lastly, the linguistic and cultural contexts of our Mandarin dataset introduce nuances in expressing politeness and refusal that differ from those in English.

To the best of our knowledge, few research has yet been conducted in classifying marketing phone call recordings, as such benchmark dataset has not been made publicly accessible. To address this research gap and facilitate future work on this complex task, we curate a new benchmark dataset, MarketCalls, which will be made available upon request. It contains 877 real-world marketing phone calls in Mandarin, each typically a minute or more in length. As one of the few publicly available, real conversational datasets for sales call analysis, and the first of its kind in Mandarin, MarketCalls provides a valuable resource for the community. It enables researchers to develop and test models on a task with greater contextual and pragmatic depth than single-utterance analysis and to investigate the generalizability of multimodal methods across different languages and cultural settings.

We propose a novel and highly effective solution, namely \textbf{M}ulti-\textbf{S}egment \textbf{M}ulti-\textbf{T}ask \textbf{F}usion \textbf{N}etwork (MSMT-FN), specifically designed to classify these phone call recordings. In marketing conversations, the explicit semantic content conveyed through texts often carries the primary signal regarding purchase intent (e.g., direct agreement or refusal). Acoustic features, such as tone, pitch, and pauses, while crucial, often serve a complementary role, refining the understanding by revealing underlying emotion, confidence, or hesitation not apparent in the transcript alone. Therefore, our network design explicitly leverages textual signals as the backbone channel, while strategically integrating complementary information from the acoustic signal using both cross-attention modules and a bottleneck fusion mechanism for effective modality fusion. Recognizing the conversational nature of the data, each audio sample is divided into multiple segments to better capture fine-grained information. We apply a bi-directional GRU (Bi-GRU) module to leverage the contextual information embedded among the segments of the same audio recording. Lastly, we embed these components within a multi-task learning framework to simultaneously address different classification granularities required by business needs and enhance MSMT-FN’s generalizability and robustness. Data augmentation is also adopted to enhance sample efficiency and reduce annotation costs.

In order to evaluate the effectiveness of MSMT-FN, we conduct comprehensive experiments using both the MarketCalls dataset and three open benchmark datasets, including CMU-MOSI, CMU-MOSEI, and MELD~\cite{poria2019meld}. Our contributions can be summarized as follows:
\begin{itemize}
    \item We create a new benchmark dataset for audio classification in marketing and make it available for the research community. The audios are in mandarin, a non-English language that can enrich research opportunities in the audio domain.
    \item We propose a novel end-to-end classification network, namely MSMT-FN, that is effective in leveraging the rich contextual signal embedded in long conversations in marketing phone calls.
    \item We demonstrate that MSMT-FN achieves strong performance on MarketCalls when compared to the state-of-the-art baseline (MMML) and generalizes well to multimodal sentiment analysis tasks on benchmark datasets.

\end{itemize}

\begin{figure*}[t] 
    \centering
    \noindent 
\includegraphics[scale=0.122]{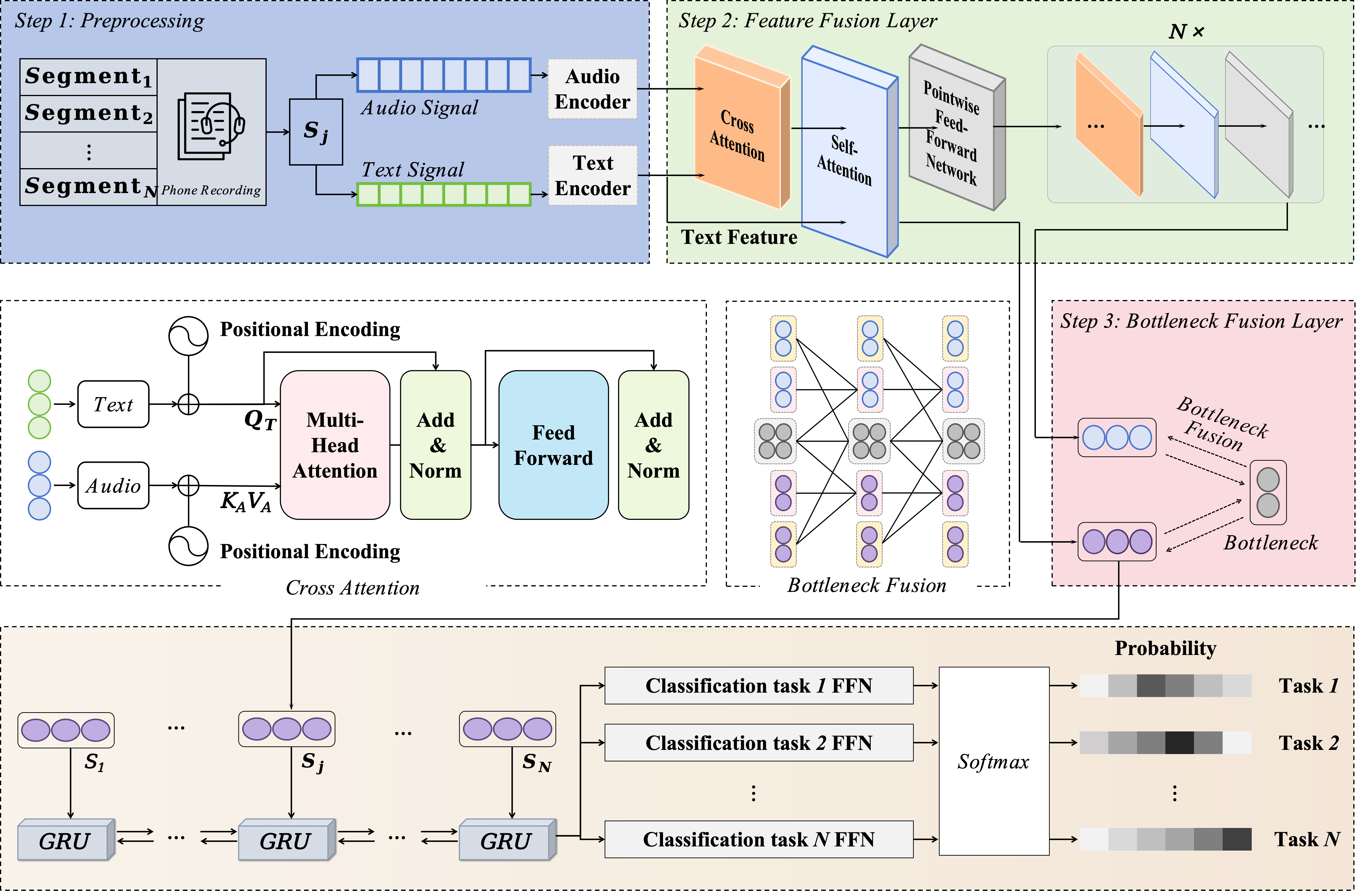}
    \caption{MSMT-FN Network Architecture. Step 1: Preprocessing. Each audio recording is broken into segments, and the audio and textual channels are extracted and encoded to obtain embeddings of the two channels. Step 2: Feature Fusion Layer. The textual channel serves as the backbone channel and is independently put through self-attention blocks; while a separate complementary channel is created by fusing the audio channel with the text channel using cross- and self-attention blocks. Step 3: Bottleneck Fusion Layer. A bottleneck fusion mechanism is adopted to more effectively fuse both channels from step 2. Finally, BiGRU is adopted under a multi-task learning framework for all segments within one audio recording to generate classification prediction for different tasks.}  
    \label{structure}
\end{figure*}

\section{Related Work}

Most existing works related to audio classification tasks concentrated in multimodal sentiment analysis and emotion recognition~\cite{joshi2022cogmen,zhang2024deep,fan2024transformer}. Marketing audio classification also falls in these broad categories, although it infers a more specific emotional propensity toward the purchasing of a given service or product being offered by a salesperson in a marketing call. To our surprise, very few high quality works exist on this important topic, particularly in the field of marketing.

In contrast to more traditional approaches using audio signal feature extraction methods~\cite{SHARMA2020107020}, most recent works applied attention-based approaches for both modality alignment~\cite{xu2020learning,goncalves2022robust} and fusion~\cite{zheng2021ctnet,yang2022multimodal,kim2023aobert,fan2024transformer,wu2024multimodal} in audio classification.

In alignment, Xu et al. proposed to use attention mechanism to learn the alignment between speech frames and text words to generate more accurate multimodal representations for emotion recognition from speech~\cite{xu2020learning}. Goncalves et al. focused on addressing the challenges in emotion recognition given non-ideal conditions for audiovisual models, including misalignment of modalities, lack of temporal modeling, and missing features due to noise or occlusions. They combined auxiliary networks in a transformer architecture using an optimized training mechanism to obtain improved accuracy and robustness~\cite{goncalves2022robust}. Yang et al. translated both the visual and audio features to textual features using BERT to tackle inferior feature qualities in visual and audio signals~\cite{yang2022multimodal}. Similarly, Kim et al. proposed the All-modalities-in-One BERT to more effectively fuse multimodal signals for sentiment analysis~\cite{kim2023aobert}.

In multimodal fusion, Zadeh et al. modeled the intra-modality and inter-modality dynamics using the proposed Tensor Fusion Network in an end-to-end approach~\cite{zadeh2017tensor}. Zheng et al. proposed to leverage transformer-based architecture to capture intra-modal and cross-modal interactions among multimodal signals~\cite{zheng2021ctnet}. Specifically, both single-modal and cross-modal transformers were adopted to unimodal and cross-modal features, and an audio-text-speaker fusion strategy was applied in fusion followed by attention mechanism to focus on important modalities. They also adopted a multi-head attention mechanism and bidirectional GRU (MHA-GRU) to extract contextual information.
Wu et al. proposed the multimodal multi-loss (MMML) fusion network~\cite{wu2024multimodal} that consists of three major components: robust feature networks was applied to extract representations from each modality; cross-modal and self-attention mechanisms were combined to capture both intra- and inter-modality dependencies in fusion; and lastly a multi-loss training strategy was adopted to further enhance fusion effectiveness. 

The above works have provided us with significant insights into feature extraction and fusion. We also adopt self-attention and cross-modal attention mechanisms in feature learning and fusion, similar to recent approaches~\cite{zheng2021ctnet,wu2024multimodal}. However, existing methods like the Multimodal Multi-Loss (MMML) fusion network~\cite{wu2024multimodal}, while powerful for general multimodal sentiment analysis on utterance-level datasets, may possess limitations when applied directly to our specific task of classifying long marketing conversations.
Firstly, such methods typically process inputs as single, often shorter, units. This might hinder their ability to effectively capture the evolving dynamics and long-range dependencies of customer intent across multiple conversational segments within a single call, a challenge our multi-segment processing coupled with Bi-GRU explicitly addresses.
Secondly, many standard fusion techniques, including attention mechanisms used in MMML, often implicitly treat modalities symmetrically or fuse them relatively early. In contrast, our approach is motivated by the nature of marketing intent classification, where text carries the primary semantic meaning. We therefore designate the textual signal as the backbone channel and design a specific fusion pathway (using cross-attention followed by bottleneck fusion) to integrate the acoustic signal as complementary information efficiently and effectively. The bottleneck fusion strategy~\cite{nagrani2021attention}, in particular, offers a structured way to manage this integration while controlling computational complexity.
Thirdly, while MMML utilizes a multi-loss strategy for potentially different aspects of sentiment, our multi-task learning framework is specifically tailored to predict customer purchase propensity at different levels of categorization granularity (e.g., two-category vs. five-category classification), directly aligning with varied business analysis requirements, and potentially improving model robustness through shared representation learning across related tasks~\cite{crawshaw2020multitask}.

By combining multi-segment contextual modeling, a text-centric complementary fusion approach via bottleneck layers, and a task-specific multi-task learning strategy, MSMT-FN aims to provide a more tailored and effective solution for classifying customer intent in complex marketing audio recordings.

\section{Method}

\subsection{Problem Formulation}

Given a recorded audio containing conversations between a salesperson and a target customer for different products and services such as dental care and cosmetic products, we want to classify the customers into different categories according to their likelihood of conversion (e.g., very positive, neutral and receptive of the call, a little impatient and negative about the call but mostly remaining being polite, explicit refusal sometimes with abusive language). These categories are manually labeled by experienced salespersons for model training purpose.

Let $A_i$ denote the audio recording from the $i$-th target customer in our dataset. To better capture nuances of each exchange term, $A_i$ can be split into conversation segments $a_j = \{a_{j,s},a_{j,c}\}$ for $j \in \{1, 2, \ldots, l_i\}$, where $l_i$ denotes the number of segments for $A_i$. Each segment $a_j$ consists of a round of conversation between the salesperson $s$ and the customer $c$. 
Let us denote the extracted representation from each audio recording $A$ as $\mathbf{X} = \mathcal{F}(A)$, the goal of the current work is to obtain a model $\mathcal{N}: \mathbf{X} \rightarrow Y$, where $Y$ denotes the purchasing propensity level.

\subsection{Data Augmentation}

To reduce costs in annotation and enhance sample efficiency, we apply several augmentation strategies that simulate real-world imperfections common in marketing phone calls, thereby improving model robustness.

\subsubsection{Audio Augmentation.} We apply a suite of augmentations to the audio signal. First, we add Gaussian noise to simulate common environmental factors, such as background noise from a customer's location (e.g., street traffics, home appliances) or call static~\cite{atmaja2016speech}. Second, to improve robustness against natural variations in human speaking rates, we apply speed perturbation, which alters audio playback speed without changing the pitch~\cite{audioaugment}. Finally, random masking is used to simulate momentary signal dropouts or packet loss common in VoIP calls, where parts of the audio become difficult for comprehension~\cite{park2019specaugment}.

\subsubsection{Text Augmentation.}
We employ Homophone Substitution, a technique particularly suited for our task. This method is designed to mimic common inaccuracies in Automatic Speech Recognition (ASR) systems, which are especially prevalent for tonal languages like Mandarin where phonetic ambiguity is high. Using the iFlytek ASR service~\cite{iFlytekASR} output, we traverse the transcript and replace each character with one of its homophones (from a pre-compiled dictionary of 8,000 characters) with a small probability (e.g., 10\%). By systematically introducing these plausible ASR errors, we force the model to learn the semantic context from surrounding words rather than over-relying on a single, potentially erroneous character. This builds a more robust language understanding that is resilient to the unavoidable imperfections of the ASR pipeline, enhancing the model's generalization capabilities.

\subsection{Audio and Text Representations}

\subsubsection{Audio Feature Extraction.}

We adopt Wav2Vec~\cite{baevski2020wav2vec} and HuBERT~\cite{HuBERT} for preprocessing the audio signal. This choice is based on past experiences, recognizing that the convolutional layers of Wav2Vec are highly effective for extracting low-level acoustic features, while HuBERT's transformer encoder is adept at capturing higher-level semantic information.

For preprocessing, audio recordings are first converted to a mono format and normalized to enhance model performance. We define a maximum segment length of 40 seconds, a decision based on observations of our phone recording samples. Segments exceeding this duration are truncated, whereas shorter segments are padded to ensure consistent input dimensions. The audio is sampled at 8,000 Hz.

Each 40-second segment, corresponding to 320,000 raw audio values (40s $\times$ 8,000 Hz), is initially processed by Wav2Vec. Its convolutional layers extract a sequence of low-level features from the raw audio. This sequence is subsequently fed into the HuBERT encoder. Leveraging its transformer architecture, HuBERT further refines these features to generate an enhanced audio representation rich in higher-level semantic content. The output of this two-stage feature extraction process is a sequence of 999 feature vectors, each 768-dimensional, for every audio segment.

Let $\mathbf{x}_{a_j} \in \mathbb{R}^{L_a \times d_a}$ denote the extracted feature vectors for segment $a_j$, where $L_a$ is the sequence length (i.e., 999 in our current case), and $d_a = 768$ is the dimension of the acoustic feature vector.
For each audio recording $A_i$, the extracted feature sequence can be represented as $\mathbf{X}_{A_i}$:
\[\mathbf{X}_{A_i} = [\mathbf{x}_{a_1}, \mathbf{x}_{a_2}, \ldots, \mathbf{x}_{a_{L_a}}]. \]

\subsubsection{Text Feature Extraction.}
The transcriptions of the conversations in each audio recording segment is tokenized and encoded using a pre-trained RoBerta\cite{RoBERTa} encoder.  
We set the maximum token length $L_t$ to 199 to ensure that the entire conversation or significant portions of it are captured within this limit. Text sequences longer than 199 are truncated, while shorter ones will be padded. Each token is embedded into a vector of dimension $d_t = 768$.

Similar to audio feature extraction, let $\mathbf{x}'_{a_j} \in \mathbb{R}^{L_t \times d_t}$ denote the extracted textual representation for the $j$th audio segment from $A_i$, we have:
\[\mathbf{X}'_{A_i} = [\mathbf{x}'_{a_1}, \mathbf{x}'_{a_2}, \ldots, \mathbf{x}'_{a_{L_t}}].\]
The feature extraction process for both audio and textual modalities is shown in Figure~\ref{structure} Step 1.

\subsection{Learning Complementary Signal from Audio}

Although our original data modality is audio, it practically contains both textual and acoustic signals. As motivated in the Section~\ref{sec:intro}, we contend that in marketing intent classification, textual content serves as the primary carrier of semantic meaning, whereas acoustic signals offer valuable supplementary information (e.g., sentiment, certainty). Our architecture reflects this by using text as a backbone and fusing audio information selectively.
For each segment $j$, we generate two pathways (as shown in Figure~\ref{structure} Step 2):

1.  \textbf{Text Backbone Pathway:} The text features $\mathbf{x}'_{a_j}$ are processed through self-attention layers to refine contextual understanding within the text itself.

2.  \textbf{Fused Complementary Pathway:} We fuse the acoustic features $\mathbf{x}_{a_j}$ with the textual features $\mathbf{x}'_{a_j}$ using cross-modal attention followed by self-attention. Specifically, we employ a cross-attention mechanism where the text representation acts as the query ($Q_T = \mathbf{x}'_{a_j}$), and the audio representation provides the keys and values ($K_A = V_A = \mathbf{x}'_{a_j}$):
    \begin{equation}
    \text{CrossAtt}(Q_T, K_A, V_A) = \text{softmax}\left( \frac{Q_T K_A^\top}{\sqrt{d_k}} \right) V_A
    \label{eq:cross_attention}
    \end{equation}
Here, $d_k$ is the dimension of the keys, and the $\text{softmax}$ function is used to normalize the attention weight. This mechanism allows the model to dynamically select acoustic features that are most pertinent to the current textual content, effectively letting the text \textbf{query} the audio for relevant complementary cues. The output from the cross-attention module is input into a self-attention module, which further improves the expressiveness of the fused feature representation. In addition, a pointwise feed-forward Network is added to further enhance the feature representation after the self-attention mechanism.
At the same time, the textual sequence is independently input into the self-attention module to enhance its representation. As is shown in Figure\ref{structure} step 2, we maintain a complementary fused channel and a separate textual channel as the input for bottleneck fusion.

\begin{table}[htbp]
\centering
    \begin{tabular}{c c c c c}
    \toprule
    \textbf{Model} & \textbf{ACC$_2$} & \textbf{ACC$_3$} & \textbf{ACC$_4$} & \textbf{ACC$_5$} \\
    \midrule
    MMML & \pmb{78.09} & 58.98 & 54.23 & 52.12 \\
    Ours & 76.60 & \pmb{63.83} & \pmb{61.70} & \pmb{60.28} \\
    \bottomrule
    \end{tabular}
    \vspace{1em}
    \caption{Experimental performances on the MarketCalls dataset. Acc is short for accuracy with the index number referring to the four different classification scenarios as described in Section~\ref{data-baselines}.}
    \label{table1}
\end{table}

\subsection{Bottleneck Fusion}
\label{sec:bottleneck_fusion}

To effectively merge the information from the text backbone pathway ($T$) and the fused complementary pathway ($T_m$) while managing computational complexity, we adapt the bottleneck fusion strategy \cite{nagrani2021attention}. The core idea is to use a small set of shared learnable \textbf{bottleneck} tokens to facilitate information exchange between the two pathways and also significantly reduce the computational complexity while maintaining information exchange through the bottleneck layers (as shown in Figure~\ref{structure}).

A set of bottleneck fusion vectors $\mathbf{T_{fsn}} = [\mathbf{T_{fsn}^1}, \mathbf{T_{fsn}^2}, \dots, \mathbf{T_{fsn}^n}]$, where $n$ denotes the number of bottleneck heads, is introduced to exchange information between the text modality features $\mathbf{T}$ and the text-audio fusion modality features $\mathbf{T_m}$. The input sequence can be expressed as:
\begin{equation}
\mathbf{T_z} = [\mathbf{T} \Vert \mathbf{T_{fsn}} \Vert \mathbf{T_m}].
\end{equation}

The information exchange and interaction between text features \( \mathbf{T} \) and text-audio fusion features \( \mathbf{T_m} \) is realized through the shared bottleneck layer \( \mathbf{T_{fsn}} \). Specifically, the text features \( \mathbf{T} \) and text-audio fusion features \( \mathbf{T_m} \) are first concatenated with the current bottleneck representation \( \mathbf{T_{fsn}} \) in the sequence dimension to form new inputs \( [\mathbf{T}, \mathbf{T_{fsn}]} \) and \( [\mathbf{T_m}, \mathbf{T_{fsn}}] \). Since the concatenation operation increases the sequence length, the system adjusts the attention mask accordingly to match the new input length to ensure that the self-attention mechanism can correctly handle the expanded sequence. Subsequently, the concatenated text channel \( [\mathbf{T}, \mathbf{T_{fsn}}] \) and text-audio fusion channel \( [\mathbf{T_m}, \mathbf{T_{fsn}}] \) are processed by the self-attention layer, as in Eq.~\ref{bottleneck1} and \ref{bottleneck2} to capture the dependencies and importance between different positions, and their respective output features are generated through the attention output layer. 

Among the generated output features, the text feature \( \mathbf{T} \) is extracted and updated from its output, and the representation of \( \mathbf{T} \) is continuously optimized using the information fused by the bottleneck layer \( \mathbf{T_{fsn}} \). At the same time, the bottleneck representation \( \mathbf{T_{fsn}} \) is also updated through the output of the text-audio fusion channel to ensure that it can effectively integrate rich information from different modalities.
Through this layer-by-layer dynamic update process, the bottleneck layer \( \mathbf{T_{fsn}} \) not only promotes the continuous optimization of the text feature \( \mathbf{T} \), but also enhances the collaborative processing capability of multimodal information, thereby significantly improving the overall expression ability and performance of the model.

We calculate the feature representation of each modality in an iterative manner as follows:
\vspace{-0.2cm}
\begin{equation} \label{bottleneck1}
[\mathbf{T^{l+1}} \Vert \mathbf{\hat{T}_{fsn}^{l+1}}] = \text{Transformer}([\mathbf{T^l} \Vert \mathbf{T_{fsn}^l}]),
\end{equation}

\vspace{-0.2cm}
\begin{equation} \label{bottleneck2}
[\mathbf{T_m^{l+1}} \Vert \mathbf{\hat{T}_{fsn,m}^{l+1}}] = \text{Transformer}([\mathbf{T_m^l} \Vert \mathbf{T_{fsn}^l}]),
\end{equation}

\vspace{-0.2cm}
\begin{equation} \label{avgt}
\mathbf{T_{fsn}^{l+1}} = \text{Avg}(\mathbf{\hat{T}_{fsn}^{l+1}},\mathbf{\hat{T}_{fsn,m}^{l+1}})
\end{equation}
 
The benefits of our proposed bottleneck fusion mechanism include the following: 
1) Shared Intermediate Representation. By introducing bottleneck representation, the bottleneck fusion mechanism establishes a shared information interaction bridge between different modalities, promoting information exchange and fusion between modalities.
2) Flexible Information Integration. The self-attention mechanism allows the model to dynamically adjust the importance of different modal information during the fusion process, capture complex dependencies, and thus achieve more flexible and effective information integration.
Through the bottleneck fusion mechanism, the model is forced to compress the features of each modality and share the most necessary information, thereby improving computational efficiency while ensuring the effect of multimodal fusion.

\begin{table}[htbp]
\centering
\renewcommand{\arraystretch}{1.3}
\makebox[\textwidth]{%
\begin{tabular}{>{\centering\arraybackslash}m{2.4cm}|>{\centering\arraybackslash}m{2.2cm}|ccccccc}
\toprule
Dataset & Model & ACC$_{\text{2Has0}}$ & F1$_{\text{Has0}}$ & ACC$_{\text{2Non0}}$ & F1$_{\text{Non0}}$ & ACC$_7$ & MAE & Corr \\
\midrule
\multirow{16}{*}{CMU-MOSI}
  & LMF     & -   & -   & 82.5 & 82.4 & 33.90 & 0.917 & 0.695 \\
  & TFN     & -   & -   & 80.8 & 80.7 & 34.90 & 0.901 & 0.698 \\
  & MFM     & -   & -   & 81.7 & 81.6 & 34.50 & 0.877 & 0.706 \\
  & MTAG    & -   & -   & 82.3 & 82.1 & 38.90 & 0.866 & 0.722 \\
  & SPC     & -   & -   & 82.8 & 82.9 & -     & -     & -     \\
  & ICCN    & -   & -   & 83.0 & 83.0 & 39.00 & 0.862 & 0.714 \\
  & MuLT    & 81.50 & 80.60 & 84.1 & 83.9 & - & 0.861 & 0.711 \\
  & MISA    & 80.79 & 80.77 & 82.10 & 82.03 & - & 0.804 & 0.764 \\
  & Self-MM & 84.00 & 84.42 & 85.98 & 85.95 & - & 0.713 & 0.798 \\
  & MAGBERT & 84.20 & 84.10 & 86.10 & 86.00 & - & 0.712 & 0.796 \\
  & MIMIM   & 84.14 & 84.00 & 86.06 & 85.98 & 46.65 & 0.700 & 0.800 \\
  & TEASEL  & 84.79 & 84.72 & 87.5  & 85    & 47.52 & 0.644 & 0.836 \\
  & UniMSE  & 85.85 & 85.83 & 86.9  & 86.42 & 48.68 & 0.691 & 0.809 \\
  & MMML    & 85.91 & 85.85 & 88.16 & 88.15 & 48.25 & \textcolor{blue}{0.6429} & \textcolor{blue}{0.838} \\
  & MMML (Con) & \pmb{87.51} & \pmb{87.45} & \textcolor{blue}{89.69} & \textcolor{blue}{89.67} & \textcolor{blue}{50.34} & \pmb{0.5831} & \pmb{0.8693} \\
  & OURS    & \textcolor{blue}{86.15} & \textcolor{blue}{86.14} & \pmb{91.21} & \pmb{91.38} & \pmb{51.02} & 0.6458 & 0.8243 \\
\midrule
\multirow{16}{*}{CMU-MOSEI}
  & LMF     & -    & -    & 82.0 & 82.1 & 48.00 & 0.623 & 0.700 \\
  & TFN     & -    & -    & 82.5 & 82.1 & 50.20 & 0.593 & 0.677 \\
  & MFM     & -    & -    & 84.4 & 84.3 & 51.30 & 0.568 & 0.703 \\
  & SPC     & -    & -    & 82.6 & 82.8 & -     & -     & -     \\
  & ICCN    & -    & -    & 84.2 & 84.2 & 51.60 & 0.565 & 0.704 \\
  & MuLT    & -    & -    & 82.5 & 82.3 & -     & 0.580 & 0.713 \\
  & MISA    & 82.59 & 82.67 & 84.23 & 83.97 & - & 0.568 & 0.717 \\
  & Self-MM & 82.81 & 82.53 & 85.17 & 85.30 & - & 0.530 & 0.765 \\
  & MAGBERT & 84.70 & 84.50 & -    & -    & - & - & - \\
  & MIMIM   & 82.24 & 82.66 & 85.97 & 85.94 & 54.24 & 0.526 & 0.772 \\
  & UniMSE  & 85.86 & 85.79 & 87.5 & 87.46 & 54.39 & 0.523 & 0.773 \\
  & MMML    & 86.32 & 86.23 & 86.73 & 86.49 & 54.95 & 0.5174 & 0.7908 \\
  & MMML (Con) & \pmb{87.24} & \pmb{87.18} & \textcolor{blue}{88.02} & \textcolor{blue}{88.15} & \pmb{55.74} & \pmb{0.4922} & \pmb{0.8137} \\
  & OURS    & \textcolor{blue}{87.16} & \textcolor{blue}{87.02} & \pmb{88.62} & \pmb{88.85} & \textcolor{blue}{55.58} & \textcolor{blue}{0.5061} & \textcolor{blue}{0.7972} \\
\bottomrule
\end{tabular}%
} 
\vspace{1em}
\caption{Experimental performances on CMU-MOSI and CMU-MOSEI. Bold numbers denote best; blue numbers denote second-best.}
\label{table2}
\end{table}

\subsection{Contextual Modeling Using Bi-directional GRU}
\label{sec:bigru}

Each audio sample is divided into conversation segments, and these segments are serving as each other's contexts. Leveraging this contextual information for classification can improve the inference accuracy. We propose to use a bi-directional GRU (Bi-GRU) layer for context modeling.

The output representations from bottleneck fusion $\mathbf{T_j}$, $j = 1,2,\ldots,l_i$ are provided as the input to the Bi-GRU layer. As is shown by Figure~\ref{structure}, each audio segment within an audio sample receives inputs from adjacent segments in both directions.
The output of the GRU layer is denoted as \(\mathbf{h}_n\), which encapsulates the final hidden states across all time steps in the input sequence.

\subsection{Multi-task Learning}

We employ a multi-task learning strategy to both accommodate business demands, as well as model generalizability and robustness.
In multi-task learning, the network backbone is shared among different related tasks as shown by Figure~\ref{structure}.
All feature representations flowing through the network backbone are input into different task-specific output layers. These layers contain independent weight parameters, and will be updated independently.
To optimize the performance of multi-task learning, each classification task computes its own cross-entropy loss, and the total loss fuction defined by Equation \ref{loss} is used to update the network parameters by backpropagation.

\vspace{-0.2cm}
\begin{equation}
    \begin{aligned} \label{loss}
        \mathcal{L}_{\text{total}} =  &   \mathcal{L}_{\text{five\_categories}} + \mathcal{L}_{\text{four\_categories}} +  \\  
        & \mathcal{L}_{\text{three\_categories}} + \mathcal{L}_{\text{two\_categories}}\\
    \end{aligned}
\end{equation}

\begin{table}[h]
\centering
\setlength{\tabcolsep}{1pt}
\renewcommand{\arraystretch}{1.05}
\begin{tabular}{c|cccccccc}
\toprule
\textbf{Model} & \textbf{ACC7} & \textbf{F17} & \textbf{surprise} & \textbf{anger} & \textbf{sadness} & \textbf{neutal} & \textbf{joy} \\
\midrule
CTNET & - & 60.5 & 52.7 & 44.6 & 32.5 & 77.4 & 56.0 \\
DF-ERC  & \pmb{68.28} & \pmb{67.03} & 60.27 & 55.50 & \pmb{43.89} & 80.17 & \pmb{65.93} \\
\midrule
OURS & 64.82 & 63.19 & \pmb{72.24} & \pmb{55.96} & 35.78 & \pmb{85.12} & 56.98 \\
\bottomrule
\end{tabular}
\vspace{1em}
\caption{Experimental performances on the MELD dataset.}
\label{table3}
\vspace{-0.5cm}
\end{table}

\section{Experiments}

\begin{table}[ht]
\centering
\caption{Statistics of the MarketCalls Dataset. The dataset initially contains 877 raw recordings.}
\label{tab:dataset_stats_booktabs} 
\begin{tabular}{l c c c}
\toprule 
\textbf{Data Split} & \textbf{Training} & \textbf{Validation} & \textbf{Test} \\
\midrule 
\multicolumn{4}{l}{\textit{Initial Raw Recordings}} \\
\quad Count & 701 & 88 & 88 \\
\midrule 
\multicolumn{4}{l}{\textit{Segmented Audio Utterances}} \\
\quad Count & 2,293 & 321 & 328 \\
\midrule
\multicolumn{4}{l}{\textit{Augmented Records (Training Set Only)}} \\
\quad Enhanced Audio & 7,041 & -- & -- \\
\quad Enhanced Text & 4,586 & -- & -- \\
\bottomrule 
\end{tabular}
\end{table}

\subsection{Data and Baselines} \label{data-baselines}
The curated dataset MarketCalls consists of 877 phone call recordings made by sales representative to potential customers in various businesses using Mandarin. These recordings are partitioned into training, validation, and test sets, as detailed in Table~\ref{tab:dataset_stats_booktabs}.

Five categories in purchase intent are created (presented in Table~\ref{tab:marketcalls_distribution}), including A - very positive, B - neutral and receptive, C - a little impatient and negative, D - explicit refusal, and E - no intent or not relevant. After data augmentation in audio and textual signals separately, we obtained a total number of 7041 and 4586 audio recording samples respectively for model training and evaluation.

In addition to the five-category task, we create three other classification tasks by combining the above five categories into the following scenarios: four-categories (A, B, C, D \& E); three-categories (A, B \& C, D \& E); and two-categories (A \& B \& C, D \& E).

\begin{table}[htbp]
    \centering
    \caption{Distribution of Labels across Domains in MarketCalls Dataset}
    \label{tab:marketcalls_distribution}
    \begin{tabular}{lccccc}
        \toprule
        \textbf{Domain}   & \textbf{Label A} & \textbf{Label B} & \textbf{Label C} & \textbf{Label D} & \textbf{Row Total} \\
        \midrule
        Dental hospital   & 82 & 128 & 223 & 57 & 490 \\
        Beauty            & 42 & 89 & 76 & 22 & 229 \\ 
        Paid courses      & 13 & 54 & 80 & 11 & 158 \\ 
        \midrule 
        \textbf{Col Total} & \textbf{137} & \textbf{271} & \textbf{379} & \textbf{90} & \textbf{877} \\ 
        \bottomrule
    \end{tabular}
\end{table}

To fully evaluate our proposed method, we also experiment with several popular open benchmark datasets, including the CMU-MOSI~\cite{zadeh2016mosi} and MOSEI datasets~\cite{bagher2018multimodal}, and the MELD dataset~\cite{poria2019meld}.
We choose recent state-of-the-art baselines in multimodal emotion recognition and sentiment analysis, including DF-ERC~\cite{li2023revisiting} and MMML~\cite{wu2024multimodal}. Other baselines are shown in various tables when appropriate.

\subsection{Hyper-parameter Tuning and Implementation}

We adopt a grid-search strategy in hyper-parameter tuning and obtain the following parameters for all experiments: learning rate is 1e-5, number of hidden layers is 4, number of accumulation steps in gradient is 4, number of bottleneck layers is 2 with 4 bottleneck nodes in each layer, number of GRU layers is 2 with 128 nodes in each layer, L2 regularization is selected and the dropout rate is 0.3.

\section{Results}

\begin{table}[htbp]
\centering
\setlength{\tabcolsep}{2pt} 
\begin{tabular}{c|cc c c c}
     \toprule
     Type & specific type & \textbf{ACC$_2$} & \textbf{ACC$_3$} & \textbf{ACC$_4$} & \textbf{ACC$_5$} \\
     \midrule 
     \multirow{3}{*}{AUG} & audio-augment & \pmb{76.60} & \pmb{63.83} & \pmb{61.70} & 60.28 \\
     & text-augment & 72.34 & 58.16 & 60.99 & \pmb{60.99} \\
     & no-augment & 75.89 & 60.02 & 58.16 & 57.46 \\
     \midrule
     \multirow{2}{*}{PREPR}& Rm-silence & 64.61 & 46.88 & 46.10 & 45.39 \\
     & Kp-silence & \pmb{76.60} & \pmb{63.83} & \pmb{61.70} & \pmb{60.28} \\
     \midrule
     \multirow{3}{*}{MODEL} & no-BF+MTL & 73.05 & 55.32 & 54.89 & 56.74 \\
     & no-MTL & 71.23 & 53.90 & 52.48 & 51.77 \\
     & full & \pmb{76.60} & \pmb{63.83} & \pmb{61.70} & \pmb{60.28} \\
     \bottomrule
\end{tabular}
\vspace{1em}
\caption{Ablation Analyses on MarketCalls dataset. 1) AUG: augmentation strategies applied on audio (audio-augment) and text (text-augment) modalities. 2) PREPR: removing (Rm-silence) or keeping (Kp-silence) silence in original audio samples. 3) MODEL: evaluation on the impacts of bottleneck fusion and multi-task learning. Bottleneck fusion and multi-task learning are removed in no-BF+MTL, while only multi-task learning is removed in no-MTL.}
\label{table4}
\vspace{-0.5cm}
\end{table}

\subsubsection{Performance Comparisons.} Table~\ref{table1}, \ref{table2} and \ref{table3} show the experimental results on the MarketCalls, CMU-MOSI and CMU-MOSEI, and MELD datasets, respectively. We can see that in the MarketCalls dataset, MSMT-FN obtains better performance in three out of four classification tasks when compared to MMML; in the CMU-MOSI and CMU-MOSEI datasets, MSMT-FN obtains better or comparable results in most evaluation metrics when compared to MMML; and in the MELD dataset, MSMT-FN obtains comparable performance when compared to DF-ERC. 

We acknowledge that no single method including the existing state-of-the-art baselines has attained consistently best performance across all evaluation metrics in all four datasets. However, our proposed network achieves comparable performances on the three open benchmark datasets in sentiment inferences, which suggests that even though our proposed method is designed mainly for customer intent classification, it can still be generalized toward a different domain.

\subsubsection{Ablation Analyses} Table~\ref{table4} shows the ablation analyses on the MarketCalls dataset with MSMT-FN. Augmenting data using the audio modality leads to best performance in all tasks. Keeping the silence portions within the audio samples results in significantly better performance than removing them. We hypothesize that this may be attributed to the fact that silent portions can contain potential signals indicative of customers' purchase intentions. Consequently, if the silent phases are removed, the model may fail to effectively capture customers' purchase intentions.
We also observe interaction effects between bottleneck fusion and multi-task learning, as the complete model achieves optimal performance. Our Bottleneck layer enables effective utilization of the text modality. However, removing only the multi-task learning component results in the poorest performance, since the model fails to adequately distinguish the boundary for purchase intent judgment, leading to suboptimal classification of multi-category tasks.

\section{Conclusion}

In this work, we propose a multi-segment multi-task fusion network that has demonstrated its effectiveness in both the MarketCalls dataset, and three open benchmark datasets. Few existing research has been conducted in this increasingly important application in the field of telemarketing, as more and more commercial activities are moving from offline to online spaces. Our contributions include both proposing a novel solution to audio intent classification, and contributing a new benchmark dataset to the field.

\bibliographystyle{splncs04}
\bibliography{main}

\end{document}